\def\td{\mbox{d}}

\documentclass[twocolumn,showpacs,preprintnumbers,amsmath,amssymb,aps,floatfix]{revtex4}

\usepackage{graphics}	
\usepackage{bm}		

\begin{document}

\title{Rules for transition rates in nonequilibrium steady states}

\author{R. M. L. Evans}
\affiliation{School of Physics and Astronomy, University of Leeds, LS2 9JT, U.K.}

\date{Accepted for publication in Physical Review Letters 20 February 2004}

\begin{abstract}
Just as transition rates in a canonical ensemble must respect the 
principle of detailed balance, constraints exist on 
transition rates in driven steady states. I derive those constraints, 
by maximum information-entropy inference, and apply 
them to the steady states of driven diffusion and a sheared lattice
fluid. The resulting ensemble can 
potentially explain nonequilibrium phase behaviour and, for steady shear,
gives rise to stress-mediated long-range interactions.
\end{abstract}


\maketitle

If we wish to design a driven stochastic model that exhibits a nonequilibrium
steady state with a given flux, how should we choose its transition rates? I
answer the question by applying Jaynes' principle of maximum 
entropy inference (MaxEnt). It might seem 
perverse to specify a macroscopic result, and then infer an equation of 
motion, since theoretical modelling usually involves the reverse procedure. 
However, for equilibrium systems, the principle of detailed balance (DB) is 
derived in just this way. The mean energy is fixed, and the equilibrium 
ensemble defined as the distribution of states respecting that constraint, 
unbiased by any other information. One then infers the properties of the 
appropriate reversible dynamics, obtaining a set of rules, DB, which demands that the ratio of 
rates for a transition and its time-reverse is given by the Boltzmann 
factor of the energy cost\cite{MCprimer}. It is commonly assumed that the same
conditions of DB should be used in nonequilibrium
models subjected to a finite through-put of flux, so that the dynamics of
local transitions is governed by the same physics as at equilibrium. However,
I shall show that those are not the transition rates predicted by MaxEnt when
a mean flux, as well as a mean energy, is specified.
The hypothesis that the phase-space paths adopted by nonequilibrium systems 
are distributed according to MaxEnt has been supported by some notable 
successes, including the recovery of linear transport theory\cite{Jaynes} 
and, more recently, the fluctuation theorem\cite{Maes} and self-organised 
criticality\cite{Dewar}. It is also a cornerstone of the 
GENERIC\cite{generic} approach to nonequilibrium kinetics.

We define a particular nonequilibrium ensemble to be the set of
phase-space paths available to a system, minimally constrained by fixing 
only the mean energy and flux on those paths. The unbiased distribution of 
paths appropriate to those constraints is given by MaxEnt. Having defined
this ensemble, we may investigate its properties without controversy. A
leap of faith is required only to hypothesise that the ensemble is a good 
description of some physical systems under realistic forcing conditions.
If so, then the method will have important implications for 
nonequilibrium phase transitions such as shear-banding\cite{banding} and 
jamming\cite{jamming}. I derive physically convincing results for 
two applications: driven diffusion, and a model of interacting particles 
under shear. 

Jaynes\cite{Jaynes} showed that Shannon's information 
entropy\cite{Shannon} $S_I=-\sum_{\Gamma}p(\Gamma)\ln p(\Gamma)$ is 
maximized by the distribution $p(\Gamma)$ of states in the equilibrium 
canonical ensemble, recovering the Boltzmann
distribution. But Jaynes also applied the method to nonequilibrium 
problems, for which $\Gamma$ represents an entire path through phase
space, spanning the duration $\tau$ of the nonequilibrium experiment in 
question. The maximization is constrained by whatever information is known 
about the paths. At equilibrium, a constant mean energy 
$\overline{E}$ is specified. To define a driven steady-state ensemble, we 
shall additionally stipulate a (possibly multi-component) mean flux 
$\overline{J}$. We shall write $p_{\tau}(X|Y)$ as the normalized 
probability for any quantity $X$ exhibited by the system during
the time interval $\tau$, subject to conditions $Y$. Thus, MaxEnt yields the 
conditional probability $p_{\tau}(\Gamma|\overline{J},\overline{E})$ that, 
over the duration $\tau$, the system takes a path $\Gamma$, 
given that the mean energy and flux have the values specified. If 
ergodicity is assumed then, for $\tau\to\infty$, $\overline{E}$ and 
$\overline{J}$ can be interpreted as time-averages. So the 
system's energy and flux are allowed to fluctuate, but their averages over 
the duration $\tau$ must have exactly the specified values on all the 
paths considered.

Our aim is to find the rate, in the driven ensemble,
\begin{equation}
\label{rate}
	\omega^{\rm driv}_{a\to b} \equiv \lim_{\Delta t\to0}
	p_{\Delta t}(a\to b | a, \overline{J}, \overline{E}) /\Delta t
\end{equation}
for a system to undergo some transition $a\to b$. This is the probability 
(per unit time) that the transition will occur within an interval 
$\Delta t$ of the current time ($t=0$), given that the current state is 
$a$, and that the energy and flux will eventually (over duration $\tau$)
have the specified time-averages. The constraints $\overline{E}$ and 
$\overline{J}$, which define the nonequilibrium ensemble, 
apply only to the duration $\tau$ as a whole; we do not constrain the 
energy and flux at each interval $\Delta t$ separately.
A conditional probability of the kind in Eq.~(\ref{rate}) can be 
manipulated by Bayes' theorem, which gives two equivalent expressions for 
the {\em joint} probability of performing the transition $a\to b$ within 
time $\Delta t$, {\em and} acquiring a mean flux $\overline{J}$ in time 
$\tau$, given the mean energy and initial state, thus:
\begin{eqnarray*}
	p_{\Delta t}(a\to b | a, \overline{J}, \overline{E})
	p_{\tau}(\overline{J}|a, \overline{E})\;= \hspace{3cm} 
	&& \\ \hspace{35mm}
	p_{\Delta t}(a\to b | a, \overline{E})
	p_{\tau}(\overline{J}|a\to b, \overline{E}). && \hspace{-35cm}
\end{eqnarray*}
Normalization implies that $p_{\tau}(\overline{J}|\ldots)$ has dimensions of reciprocal flux.
Substituting into Eq.~(\ref{rate}) and recognising that quantities for 
which only the mean energy is constrained belong to the equilibrium 
ensemble, yields
\begin{equation}
\label{relation}
	\omega^{\rm driv}_{a\to b} = \omega^{\rm eq}_{a\to b} \, \eta_{ab}.
\end{equation}
Equation (\ref{relation}) confirms, as asserted above, that the rate of a 
transition $a\to b$ in the driven ensemble is not equal to the rate at 
equilibrium, but is enhanced by a factor 
\begin{equation}
\label{enhancement}
	\eta_{ab} \equiv \lim_{\tau\to\infty}
	\frac{ p_{\tau}^{\rm eq}(\overline{J}|a\to b) }
	{ p_{\tau}^{\rm eq}(\overline{J}|a) }.
\end{equation}
This factor is known in principle, as it is a property of the corresponding 
system at equilibrium, not in the driven ensemble. The denominator in 
Eq.~(\ref{enhancement}) is the probability that a system, starting in the 
current state $a$, will exhibit an average flux $\overline{J}$ over the 
duration $\tau$ if it is governed by the equilibrium transition rates. Of 
course it is exceedingly unlikely for an equilibrium system to 
spontaneously perform sustained flow, so the denominator is 
infinitesimal. The numerator in Eq.~(\ref{enhancement}) measures the 
infinitesimal probability of that same flux at equilibrium, given that the 
dynamics begins with a transition to state $b$. We shall see that 
Eqs.~(\ref{relation}, \ref{enhancement}) make intuitive sense in some
examples below.

The above derivation exploits the fact that the driven ensemble is a 
sub-set of the equilibrium ensemble (albeit in the extreme tail of the 
flux distribution), since it is defined by one extra constraint. 
But the `sub-ensemble dynamics' (SED)
[Eqs.~(\ref{relation}, \ref{enhancement})] should not be mistaken for a 
near-equilibrium approximation since the sub-set of paths has 
properties very different from the equilibrium set. 
Nevertheless, an equilibrium Markov process remains Markovian under 
SED. For many transitions, that contribute no flux and do not alter the 
future likelihood of flux, $\eta_{ab}=1$ so Eq.~(\ref{relation}) says 
the rate is equal in the driven and equilibrium ensembles, as often
assumed. Two types of transition are boosted in the driven 
ensemble: (A)~a transition that carries a positive flux in the direction of 
$\overline{J}$, (B)~a transition to a state that is more amenable to 
subsequent flux-carrying transitions.

Many choices of prior rates $\omega^{\rm eq}_{a\to b}$ are possible, either fully implementing Newton's 
laws, or embodying approximate (e.g.~Brownian) dynamics. Consider the exact 
Newtonian evolution of the particles of a fluid element surrounded by a 
reservoir of more fluid. Unphysical transitions, e.g.~violating momentum 
conservation for internal degrees of freedom, have zero prior rate, so 
Eq.~(\ref{relation}) also forbids such transitions in the driven case, 
e.g.~under shear flow. Thus, the scheme respects Newton's laws and is 
consistent with the Liouville equation. The reservoir introduces randomness 
into the dynamics by coupling to particles at the surface of the fluid 
element. SED provides the unbiased description of the reservoir's influence.

We now study two examples with Brownian prior dynamics. The first, a driven 
Brownian ideal gas, is simple enough for exact calculation, but exhibits 
only
`type A' transitions. More of the physical richness of SED will appear 
in the second example, a complex system under shear that demonstrates both 
types of enhancement.

Let us find the equation of motion for the particles of a Brownian ideal gas 
with a drift velocity $v$. The problem
de-couples for each component of each 
particle's displacement. The non-trivial part is the component in the flux 
direction. We must assume prior knowledge of the motion at equilibrium, for 
which each coordinate $x(t)$ performs an unbiased random walk according 
to the Langevin equation,
\mbox{$\zeta \td x/\td t = \xi(t)$}
with $\zeta$ a friction constant, and $\xi$ a delta-correlated noise 
function\cite{Langevin}. We introduce a time step $\Delta t$ (that will 
eventually be taken to zero), so that the thermal noise $\xi$ is drawn 
from a well-behaved Gaussian distribution\cite{Langevin} and the Langevin 
equation relates this stochastic variable $\xi$ to a step $x\to x+\Delta x$ 
such that $\Delta x = \xi\, \Delta t/\zeta$.
This equilibrium dynamics dictates that each Brownian particle follows a 
path with steps drawn from the distribution
\begin{equation}
\label{eqprob}
	p_{\Delta t}^{\rm eq}(\Delta x) = G(\Delta x,\Delta t)
	= \frac{1}{\sqrt{4\pi D \Delta t}} 
	\exp \left(\frac{-\Delta x^2}{4D\Delta t}\right)
\end{equation}
which corresponds (per unit time) to a transition rate (per unit distance) 
$\omega^{\rm eq}_{x\to x+\Delta x}$ in the notation of Eq.~(\ref{relation}). 
Here, $G(x,t)$ is the Green function for free diffusion, and the 
diffusion coefficient is given by the Einstein relation $D=k_B T/\zeta$. 
Now, from the set of all equilibrium paths, we extract the sub-set 
exhibiting the required flux $v$, 
by introducing {\em a posteriori} the constraint
$x(\tau) = v\tau \equiv x_0$ \cite{comment}.
On those paths, the rate of a step $\Delta x$ from position $x$ at 
time $t$ is enhanced by
\begin{equation}
\label{diffprob}
	\eta_{x, x+\Delta x} =
	\frac{ G(x_0-x-\Delta x, \tau-t-\Delta t) }{ G(x_0-x, \tau-t) }
\end{equation}
[from Eq.~(\ref{enhancement})], since the probability of an equilibrium
particle achieving the required displacement in the remaining time is 
given by the Green function for free diffusion. Substituting 
Eq.~(\ref{diffprob}) into Eq.~(\ref{relation}) in the limits $\tau\gg t$ 
and $|x_0|\gg|x|$ yields
\begin{eqnarray}
	p_{\Delta t}^{\rm driv}(\Delta x)
	&\to& p_{\Delta t}^{\rm eq}(\Delta x) \,
	\left[ 1+\Delta x\,\frac{v}{2D} 
	- \Delta t\,\frac{v^2}{4D} \right] \nonumber \\
	&\to& p_{\Delta t}^{\rm eq}(\Delta x-v\,\Delta t) 
\label{drivenprob}
\end{eqnarray}
where the second line follows by inspection of Eq.~(\ref{eqprob}). 
Drawing $\Delta x$ from $p_{\Delta t}^{\rm eq}$ in Eq.~(\ref{eqprob}) 
yields the equilibrium dynamics. So Eq.~(\ref{drivenprob}) specifies that 
substituting \mbox{$\Delta x-v\,\Delta t$} for $\Delta x$ in the 
equilibrium equation of motion will yield the dynamics of the driven 
Brownian ideal gas. Making that substitution, with $\Delta t\to 0$, gives
\begin{equation}
\label{drivendiffusion}
		\frac{\td x}{\td t} = v + \xi(t)/\zeta
\end{equation}
where $\xi$ is the usual Gaussian white noise. This is the equation of 
motion for driven diffusion that one could easily have written down. But 
it was not conjectured; it was generated from the Langevin equation for 
free diffusion, by the sub-ensemble method. In fact, one could write any 
number of stochastic equations that yield the net drift $v\tau$ over the 
fixed duration $\tau$, e.g.~with some temporal correlations or an 
additional oscillatory forcing term that integrates to zero. But 
Eq.~(\ref{relation}) specifies a unique solution. Any equation of motion 
other than the simple one specified by these dynamics would violate 
the MaxEnt hypothesis, indicating that it introduces unwarranted new 
information about the paths, additional to the prior dynamics and 
posterior constraint.

Let us consider a second example, demonstrating that rates respecting DB are not generally correct for driven systems. 
In a simple 2D model of Brownian particles under continuous shear, a 
triangular lattice has some fraction of its sites occupied by monomers with 
nearest-neighbour interactions. An average velocity difference between the 
top and bottom boundaries is established by stochastically selecting a 
horizontal layer $l$ between two rows of the lattice, and shifting all of the 
system above this layer to the right by one lattice spacing. If the layers 
are selected with equal probability, these discrete shear transitions will 
result in a uniform shear rate when observed on large length- and 
time-scales. In addition to the shear steps, local dynamics consists of 
choosing a pair $s$ of neighbouring sites at random and swapping them, with 
a rate $\omega_s$. If one site is occupied by a monomer and the other empty, the transition causes the monomer to hop, and might result in an energy change by making or breaking bonds. Repeating the same swap $s$ recovers the original 
configuration. In the absence of shear, transition rates 
respecting DB, e.g.~`heat-bath 
dynamics'\cite{MCprimer},  will correctly generate all static correlations.
With shear applied to our model, let us initially violate 
Eq.~(\ref{relation}), and assume that the same equilibrium dynamics is 
chosen for the local swap transitions.

\begin{figure}[!t]
  \resizebox{68mm}{!}{\includegraphics{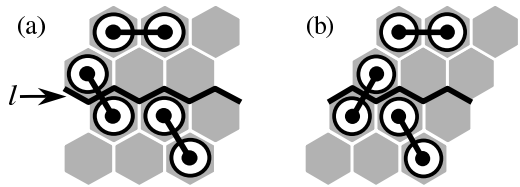}}
  \caption{\label{swivel}
	(a) Before, (b) after a shear step on layer $l$, displacing 
	everything above it to the right, thus rotating a dimer clockwise, 
	and making new contacts between monomers.}
\end{figure}

The model as defined would settle into a driven steady state in the 
long-time limit and, by tuning the interactions, could be made to 
exhibit nonequilibrium phase transitions and amusing mesophase 
structures. But it is unphysical, as becomes apparent with a 
particular choice of interactions. Let the monomers be coupled in 
pairs by bonds of infinite strength, to form dimers. 
A shear transition rotates some dimers (Fig.~\ref{swivel}). Any dimer
in a north-east south-west orientation prevents shear on the layer that it 
straddles, e.g.~in Fig.~\ref{swivel}b further shear on the 
same layer is disallowed by a dimer that would be 
broken by the transition. In the thermodynamic limit, the chance of 
finding a lattice layer that is not blocked by at least one such 
adversely oriented dimer vanishes. The model cannot be driven to shear.

It is unreasonable for a dimer to wander through configurations 
under equilibrium-like dynamics, unaware that it is blocking a macroscopic 
flux. Clearly such dynamics will not work for rigid Brownian dimers.
Instead, DB should be violated, even at the local
level, so that adversely oriented dimers are pushed out of the way 
of the applied flux. Even a dimer of finite strength should prefer
re-orientation to shear-induced dissociation. The rate of
re-orientation is prescribed by SED.

Consider first how SED treats shear steps. A shear step is 
a `type A' transition as it contributes a quantum of shear flux. In an 
{\em equilibrium} model with both forward and reverse shear steps, 
the probability of accumulating a large net 
shear by time $t=\tau$ is tiny. If a forward shear transition $a\to b$
takes place before $t=\Delta t$, then one step fewer is subsequently 
required to attain the desired shear, so the numerator of 
Eq.~(\ref{enhancement}) is larger than the denominator. Thus 
Eq.~(\ref{relation}) correctly prescribes a higher rate for forward shear 
steps in the driven ensemble than at equilibrium. Similarly, reverse steps 
are suppressed. As shear steps on 
different layers $l$ (see Fig.~\ref{swivel}) contribute equal flux, 
they have approximately equal enhancement factors. But their rates 
$\omega^{\rm driv}_l$ are not necessarily equal (yielding affine shear) 
since, by DB, $\omega^{\rm eq}_l$ in Eq.~(\ref{relation}) 
depends on the energy cost. Hence the driven shear steps are concentrated 
on the softest layers, e.g.~where fewest attractive neighbours will be 
separated. Rather than imposing affine shear, the driven ensemble, with its
weak constraint fixing only the total mean flux, allows authentic 
inhomogeneous flow.

\begin{figure}[!t]
  \resizebox{45mm}{!}{\includegraphics{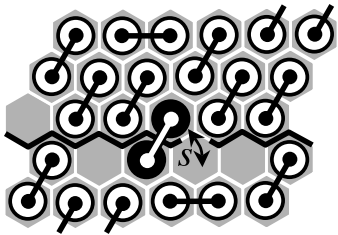}}
  \caption{\label{stressconc}
	A configuration that cannot admit a forward shear step. After swap 
	$s$, the highlighted layer can shear.}
\end{figure}

As well as prescribing the rate of shear steps, Eq.~(\ref{relation}) also 
governs the local site-swapping dynamics.
The site-swapping transition $s$ 
indicated in Fig.~\ref{stressconc} is not of `type A', as it contributes no 
shear flux, but we shall see that its rate is boosted by SED 
as it is a `type B' transition. In Fig.~\ref{stressconc}, forward 
shear is blocked on every lattice layer by adversely-oriented dimers. So 
there is no chance of immediate shear and, of course, little hope of the 
desired net flux over the duration $\tau$ of an equilibrium experiment. 
Hence the denominator of Eq.~(\ref{enhancement}) is very small. The 
numerator asks for the likelihood of that same shear flux if swap $s$ 
(represented by $a\to b$) is 
first performed. This would rotate an offending dimer, allowing forward 
shear on the layer shown in bold. A shear step is not guaranteed to follow 
swap $s$, but its likelihood is greatly increased by the swap. Further 
shear could follow, until eventually some dimers re-block 
the layer, returning to a configuration statistically similar to 
Fig.~\ref{stressconc}. Hence, if $s$ is performed, the 
probability of achieving the desired net flux at equilibrium, 
although small, is many times larger than the denominator of 
Eq.~(\ref{enhancement}). So SED greatly boosts the 
rate for swap $s$ in Fig.~\ref{stressconc}, and the highlighted dimer is 
quickly `pushed out of the way'. Similarly, if transition $s$ were blocked 
by another dimer it too would be moved by SED, and
so on, with correlated chains of events enabling the stipulated mean flux to 
be realised.

It is startling that SED has generated long-range 
couplings governing the local swaps, whereas their prior 
equilibrium dynamics depends only on nearest-neighbour interactions. 
Because the dimer highlighted in Fig.~\ref{stressconc} is alone in 
blocking its layer of the lattice, it is quickly rotated under 
applied shear. It feels that there are no other soft planes in the whole 
lattice that could yield. If there were, the dimer would not feel such an 
imperative to move, as the enhancement factor would be smaller. We 
interpret that stress is concentrated at this point. The physics 
of long-range stress-mediated interactions has arisen naturally.

Unlike the example of the Brownian gas, the Green function for the 
dimers is unknown, so an exact calculation is not presented. The problem is 
that evaluation of Eq.~(\ref{enhancement}) appears to 
require clairvoyance of the full consequences of a proposed transition. 
However, those consequences need only be forecast for a finite time into 
the future, exceeding any correlation time, following which the steady 
state can be assumed. The problem is generally tractable in terms of a 
cluster expansion\cite{unpub}. We can speculate that the limit in which such 
an expansion breaks down (because many correlated particle movements are 
required for any finite flux probability) might be identified with a jamming
transition\cite{jamming}, just as break-down of the virial expansion 
accompanies a critical point.

The method outlined here (detailed elsewhere\cite{unpub}) 
puts the simulation of driven steady 
states on a closer footing to equilibrium numerics, for which any 
DB-respecting algorithm yields the Gibbs ensemble. By 
contrast, sheared fluids have hitherto required microscopically accurate 
simulations\cite{DEvans}, with the associated processing overhead and 
thermostatting issues\cite{thermostat}.
This scheme is not the only route to nonequilibrium 
transition rates. Models may be defined that violate Eq.~(\ref{relation}), 
but one should then be aware that extraneous information has been 
introduced, not present in the prior dynamics and macroscopic 
observables. In this respect the rules set out here have the same status 
as the principle of DB. 

The driven ensemble has been defined here by fixing the time-averaged flux
on {\em each} path, analogous to fixing energy in an equilibrium 
microcanonical ensemble. In the limit 
$\tau\to\infty$ (analogous to the thermodynamic limit), identical results 
follow from an alternative ensemble, in which paths are weighted 
exponentially by their flux, analogous to the canonical ensemble. Details 
will appear in a longer paper\cite{unpub}, but note that 
simply boosting a transition exponentially by the immediate flux that it 
carries would neglect the subtle non-mean-field time correlations of SED.

I have studied the physical implications of MaxEnt for driven steady 
states and shown how to implement the resulting dynamics. The approach is 
unique in deriving dynamical rules for a driven system under the stochastic 
influence of a reservoir, without requiring any approximate coarse-graining or near-equilibrium assumption. The rules have yielded a rich variety of correct physics for 
driven diffusion and
Brownian dimers under shear. Many quiescent systems 
are well approximated by the laws of canonical equilibrium, but exceptions 
include glasses, granular media, and some cellular automata. Similarly, not 
every nonequilibrium steady state will respect the conditions presented 
here, but those that do are expected to form a large and significant class.

Thanks go to A. D. Bruce, M. E. Cates, R. A. Blythe, P. D. Olmsted, 
T. C. B. McLeish,
A. E. Likhtman, and S. M. Fielding for informative 
discussions.
RMLE is a Royal Society University Research Fellow.


\begin{thebibliography}{99}

\bibitem{MCprimer} M. A. Novotny, cond-mat/0109182 (2001); K. A. Fichthorn and W. H. Weinberg, J. Chem.\ Phys.~{\bf 95}, 1090 (1991).

\bibitem{Jaynes} E. T. Jaynes, Phys.\ Rev.~{\bf 106}, 620 (1957); {\bf 108}, 171 (1957); E. T. Jaynes, in ``Maximum entropy formalism", Eds R. D. Levine and M. Tribus (MIT Press, Cambridge Massachusetts 1979);

\bibitem{Maes} C. Maes, J. Stat.\ Phys.~{\bf 95}, 367 (1999).

\bibitem{Dewar} R. Dewar, J. Phys.\ A {\bf 36}, 631 (2003).

\bibitem{generic} M. Grmela and H. C. \"{O}ttinger, Phys.\ Rev.~E {\bf 56}, 6620 (1997).

\bibitem{banding} P. D. Olmsted, Europhys.~Lett.~{\bf 48}, 339 (1999). See also S. Butler and P. Harrowell, Nature {\bf 415}, 1008 (2002).

\bibitem{jamming}J. R. Melrose and R. C. Ball, Europhys.\ Lett.~{\bf 32} 535, (1995); M. E. Cates, J. P. Wittmer, J.-P. Bouchaud and P. Claudin, Phys.\ Rev.\ Lett.~{\bf 81}, 1841 (1998).

\bibitem{Shannon} C. E. Shannon, Bell System Tech.~J. {\bf 27}, 379 (1948); 623.

\bibitem{Langevin} P. M. Chaikin and T. C. Lubensky, ``Principles of condensed matter physics" (Cambridge University Press, 1997).

\bibitem{comment} Strictly, the constraint applies to the gas on average, not
each particle separately. But, in the limit $\tau\to\infty$,
fluctuations about the mean can be neglected.

\bibitem{unpub} R. M. L. Evans, in preparation.

\bibitem{DEvans} D. J. Evans, E. G. D. Cohen and G. P. Morriss, Phys.~Rev.~Lett.~{\bf 71}, 2401, (1993); 3616; D. J. Evans and G. P. Morriss, ``Statistical mechanics of nonequilibrium liquids" (Academic, London 1990).

\bibitem{thermostat} A. Baranyai and P. Cummings, Mol.\ Phys.~{\bf 90}, 35 (1997).

\end{thebibliography}
\end{document}